\documentclass[11pt]{article}
\pdfoutput=1
\usepackage{amsmath,amssymb,color,cite}
\usepackage{slashed}
\usepackage{graphicx}
\usepackage{amsfonts}
\usepackage{enumerate}
\usepackage{hyperref}
\usepackage{bbm}
\usepackage{nicefrac}
\usepackage[all]{xy}
\usepackage{graphicx}
\usepackage{booktabs}
\usepackage{epstopdf}
\usepackage{simplewick}
\usepackage{amsfonts}

\newcommand{\hoch}[1]{$\, ^{#1}$}

\makeatletter
\@addtoreset{equation}{section}
\makeatother

\newcommand{\imineq}[2]{\vcenter{\hbox{\includegraphics[height=#2ex]{#1}}}}

\newcommand{\be}{\begin{equation}}
\newcommand{\ee}{\end{equation}}
\newcommand{\bea}{\setlength\arraycolsep{2pt} \begin{eqnarray}}
\newcommand{\eea}{\end{eqnarray}}
\newcommand{\nn}{\nonumber}

\def\0{{\sst{(0)}}}
\def\1{{\sst{(1)}}}
\def\2{{\sst{(2)}}}
\def\3{{\sst{(3)}}}
\def\4{{\sst{(4)}}}
\def\5{{\sst{(5)}}}
\def\6{{\sst{(6)}}}
\def\7{{\sst{(7)}}}
\def\8{{\sst{(8)}}}
\def\sst#1{{\scriptscriptstyle #1}}

\begin{document}
\begin{flushright}
\hfill{ \
\ \ \ \ }
\end{flushright}

\vspace{25pt}
\begin{center}
{\Large $\phi^3$ theory with $F_4$ flavor symmetry in $6-2\epsilon$ dimensions: 3-loop renormalization and conformal bootstrap {\bf }
}

\vspace{30pt}

{\Large
Yi Pang\hoch{1}, Junchen Rong\hoch{2} and Ning Su\hoch{3}
}

\vspace{10pt}
\hoch{1} {\it Max-Planck-Insitut f\"{u}r Gravitationsphysik (Albert-Einstein-Institut)\\
Am M\"{u}hlenberg 1, DE-14476 Potsdam, Germany}

\hoch{2} {\it Fields, Gravity \& Strings, Center for Theoretical Physics of the Universe,
Institute for Basic Sciences, Daejeon 305-811, Korea
}\\

\hoch{3} {\it CAS Key Laboratory of Theoretical Physics, Institute of Theoretical Physics, Chinese Academy of Sciences, Beijing 100190, China
}
\vspace{10pt}

\vspace{20pt}

\underline{ABSTRACT}
\end{center}
\vspace{15pt}

We consider $\phi^3$ theory in $6-2\epsilon$ with $F_4$ global symmetry. The beta function is calculated up to 3 loops, and a stable unitary IR fixed point is observed. The anomalous dimensions of operators quadratic or cubic in $\phi$ are also computed. We then employ conformal bootstrap technique to study the fixed point predicted from the perturbative approach. For each putative scaling dimension of $\phi$ ($\Delta_{\phi})$, we obtain the corresponding upper bound on the scaling dimension of the second lowest scalar primary in the ${\mathbf{26}}$ representation $(\Delta^{\rm 2nd}_{{\mathbf{26}}})$ which appears in the OPE of $\phi\times\phi$. In $D=5.95$, we observe a sharp peak on the upper bound curve located at $\Delta_{\phi}$ equal to the value predicted by the 3-loop computation. In $D=5$, we observe a weak kink on the upper bound curve at $(\Delta_{\phi},\Delta^{\rm 2nd}_{{\mathbf{26}}})$=$(1.6,4)$.

\thispagestyle{empty}

\pagebreak
\voffset=-40pt
\setcounter{page}{1}



\section{Introduction}
 Conformal field theory (CFT) describing interesting infrared (IR) physics usually arises as the fixed point of renormalization group flow. A useful perturbative tool to study such kind of fixed point is the $\epsilon$-expansion, which has been applied to explore the IR fixed point of quartic scalar theory in $D=4-\epsilon$ dimensions, including $D=3$ Ising model \cite{Wilson:1971dc} and critical O($N$) vector model (see \cite{Moshe:2003xn} for a comprehensive review). In $4< D< 6$, the quartic scalar interaction becomes irrelevant and the renormalization group flow can instead be triggered via a cubic scalar interaction. The simplest $\phi^3$ theory in $6-2\epsilon$ has been considered long time ago \cite{Macfarlane:1974vp, Ma:1975vn, Fisher:1978pf}, with the Lagrangian $\mathcal{L}=\frac{1}{2}(\partial\phi)^2+\frac{1}{6}g \phi^3$. In \cite{Fisher:1978pf}, it was shown that the 1-loop beta function has a non-unitary IR fixed point with imaginary coupling constant $g$ for $D<6$. Continuation of this fixed point to $D=2$ describes the Yang-Lee edge singularity \cite{Yang:1952be,Lee:1952ig} in the Ising model (this is the (2,5) minimal model \cite{Belavin:1984vu,Cardy:1985yy} with negative central charge).
 Recently there has been a revival of interests to the renormalization of quantum field theory with $\phi^3$ interaction in $D=6-2\epsilon$ \footnote{The coefficient 2 in front of $\epsilon$ is our choice of convention.} \cite{Fei:2014yja,Grinstein:2014xba,Fei:2014xta,Fei:2015kta,Grinstein:2015ina,Gracey:2015tta,Stergiou:2016uqq}, motivated by studying $a$-theorem in $D> 4$ or higher spin holography. In particular, the Lagrangian
\be
\mathcal{L}=\frac{1}{2}(\partial\phi^i)^2+\frac{1}{2}(\partial\sigma)^2+\frac{g_1}{2}\sigma\phi^i\phi^i+\frac{g_2}{6}\sigma^3
\ee
was utilized in \cite{Fei:2014yja,Fei:2014xta} to investigate the $D=5$ critical O($N$) vector model \footnote{See \cite{Herbut:2015zqa} for a study of tensorial O(N) model.}. An interesting phenomenon originally noted in \cite{Ma:1975vn} and recently rediscovered in \cite{Fei:2014yja}  is that there exists a critical value for $N$, denoted as $N_{\rm crit}$, above which a stable, unitary fixed point was found in $6-2\epsilon$ dimension. One-loop renormalization suggests that $N_{\rm crit}\approx 1038$ \cite{Ma:1975vn, Fei:2014yja}. Later, a 3-loop computation implies a much smaller $N_{\rm crit}$ \cite{Fei:2014xta}.

As a non-perturbative approach to CFT, conformal bootstrap dates back to the work of Alexander Polyakov \cite{Polyakov:1974gs} and also the work of Sergio Ferrara, Raoul Gatto and Aurelio Grillo \cite{Ferrara:1973yt} in the 1970s . Its later application to two dimensional conformal field theories led to the famous work of  Alexander Belavin, Alexander Polyakov and Alexander Zamolodchikov \cite{Belavin:1984vu}, which classified $D=2$ minimal models.  In $D>2$, a significant progress was made by  \cite{Rattazzi:2008pe}. Since then, conformal bootstrap has been applied to $D=3$ Ising model\cite{ElShowk:2012ht,Kos:2014bka}, O($N$) vector models\cite{Kos:2013tga,Kos:2016ysd}, Gross-Neveu(-Yukawa) models\cite{Iliesiu:2015qra} and other CFTs with or without supersymmetry \cite{Vichi:2011ux,Beem:2013qxa,Beem:2014zpa,Bobev:2015jxa,Chester:2015lej,Chester:2014fya,Beem:2015aoa,Bobev:2015vsa,Lin:2015wcg,Bashkirov:2013vya}. As a powerful non-perturbative method, for instance, conformal bootstrap has improved the precision of critical exponents in $D=3$ Ising model by two orders of magnitude, compared to the Monte-Carlo simulations \cite{Kos:2016ysd}. In $D=5$, attempts to bound the value of $N_{\rm crit}$ for critical O($N$) vector model has been carried out through conformal bootstrap approach \footnote{See \cite{Mati:2014xma,Mati:2016wjn,Eichhorn:2016hdi,Kamikado:2016dvw} for the study of O($N$) vector model using non-perturbative method "Functional Renormalization Group".} \cite{Bae:2014hia,Chester:2014gqa,Li:2016wdp}. In \cite{Chester:2014gqa}, using the single correlator bootstrap, it was observed that a kink which exists for large enough $N$ ceases to exist when $15<N_{\rm crit}<22$, under a reasonable assumption on the scaling dimension of the second lowest O$(N)$ singlet scalar primary. On the other hand, the mixed correlators bootstrap seems to suggest that $N_{\rm crit}>100$ \cite{Li:2016wdp}.

In this work, we explore the possibility of having a CFT in five dimensions with $F_4$ global symmetry. The exceptional Lie group $F_4$ known as the compact real form of Lie algebra $\mathbf{f_4}$, is also the isometry group of the octonionic projective plane $\mathbf{OP}^2$ \cite{Baez:2001dm}. It admits a rank-2 and a rank-3 irreducible symmetric invariant tensors denoted by $\delta_{ij}$ and $d_{ijk}$, where the index transforms as the $\mathbf{26}$ of $F_4$. The simplest interacting $F_4$ theory can be written as a scalar theory with a cubic self-interaction
\be\label{F4Lagragian}
\mathcal{L}=\frac{1}{2}\delta_{ij}(\partial_{\mu}\phi^i)(\partial^{\mu}\phi^j)+\frac{g}{6}d_{ijk}\phi^i\phi^j\phi^k.
\ee
The cubic interaction is relevant in $6-2\epsilon$ dimensions and may drive the theory to a nontrivial IR fixed point. From the 3-loop renormalization of the coupling constant, we indeed observe a stable IR fixed point in $D=6-2\epsilon$. We then employ conformal bootstrap technique to probe such a fixed point in $D=5.95$ and $D=5$ \footnote{ It was shown  in\cite{Hogervorst:2014rta} that the O($N$) vector model in non-integer dimensions was non-unitary. However, conformal bootstrap approach is still applicable in non-integer dimensions \cite{Bae:2014hia,Chester:2014gqa,Li:2016wdp} and leads to reasonable results which can be compared with those derived from $\epsilon$-expansion. We expect the same here for $F_4$ invariant $\phi^3$ theory.}. We observe that in $D=5.95$ the boundary of the allowed region in the $(\Delta_{\phi},\Delta_{\mathbf{26}}^{\rm 2nd})$ plane exhibits a sharp peak exactly at the value of $\Delta_{\phi}$ obtained from the Pad\'{e}$_{[2,1]}$ resummed 3-loop results. In $D=5$, a weak kink is observed near the 3-loop results. The appearance of the kink has to do with fact that when the anomalous dimension of $\phi^i$ is small, the second lowest scalar primary in $\mathbf{26}$ is approximately given by $d_{ijk}\phi^j\phi^k$ with dimension $2\Delta_{\phi}$. However, when the anomalous dimension $\phi^i$ is large enough, the theory acquires notable deviation from the free theory. In the interacting theory, the operator $d_{ijk}\phi^j\phi^k$ becomes a conformal descendant of $\phi^i$. The new second lowest scalar primary in $\mathbf{26}$ should have much higher dimension that $2\Delta_{\phi}$, yielding a sudden change in $\Delta_{\mathbf{26}}^{\rm 2nd}$ .

This paper is organized as follows. In Section \ref{renormalization}, we present the renormalization of the theory \eqref{F4Lagragian} in $D=6-2\epsilon$ using $\epsilon$-expansion. In particular, we compute the anomalous dimensions of operators such as $\phi^i$, $\phi^i\phi^i$ and  $d_{ijk}\phi^i\phi^j\phi^k$ up to ${\cal O}(\epsilon^3)$. We also calculate the anomalous dimensions of other $\phi^2$ operators in $\mathbf{26}$ and $\mathbf{324}$ representations at 1-loop level. In Section \ref{conformalbootstrap}, we derive the set of crossing equations for a CFT with $F_4$ global symmetry and then apply it to study the fixed points predicted by the loop calculations in $D=5.95$ and $D=5$, using numerical conformal bootstrap. We discuss future extensions in Section \ref{discussion}.

\subsection{3-Loop Renormalization of generic $\phi^3$ theory in $6-2\epsilon$ Dimensions}\label{general3loop}
The 3-loop renormalization of generic $\phi^3$ theory in $D=6-2\epsilon$ was studied long time ago by \cite{deAlcantaraBonfim:1980pe,deAlcantaraBonfim:1981sy} using the modified minimal subtraction (MS)
scheme. Recently \cite{Gracey:2015tta} has extended the results to 4 loops. Here we only utilize the 3-loop results. The 3-loop beta function is given by \footnote{Throughout this paper, we mainly follow the convention used in \cite{Gracey:2015tta}, except that the sign of $g^2$ has been reversed. Compared with \cite{Fei:2014yja,Fei:2014xta}, there is a factor of 2 difference in the definition of $\beta(g)$, and the numerical factor $\text{Area}(S_5)/(2\pi)^6=(4\pi)^3$ has been included in $g^2$.}
\bea
&&\beta(g)=-\frac{\epsilon}{2}g+ \frac{ T_2 - 4T_3}{8}g^3
+ \frac{ 66T_2T_3- 11T_2^2 - 108T_3^2 - 72T_5}{288}g^5\nn\\
&&+\Big( 821T_2^3 - 6078T_2^2T_3 + 12564T_2T_3^2 - 2592T_2T_5\zeta(3) + 9288T_2T_5
+ 11664T_3^3\\
&& + 51840T_3T_5\zeta(3) - 61344T_3T_5 - 20736T_{71} - 62208T_{72}\zeta(3)
+ 20736T_{72}\Big)\frac{g^7}{41472} +O(g^9)\,.\nn
\label{betafunction}
\eea
The 3-loop anomalous dimension of $\phi$ takes the form
\bea\label{anadimension}
&&\gamma_{\phi}=  \frac{T_2}{12}g^2
+ \frac{ 24 T_2 T_3- 11 T_2^2}{432}g^4 \nn\\&&
+\Big(  821T_2^3 - 3222T_2^2T_3 + 3060T_2T_3^2 - 2592T_2T_5\zeta(3)
+ 4536T_2T_5\Big)\frac{g^6 }{62208}+O(g^8)\,,
\eea
and the anomalous dimension of operator $\mathcal{O}\sim\phi^i\phi^i$  (for simplicity, from now on we will denote the $F_4$ singlet $\phi^2$ operator by $\phi^2\in \mathbf{1}$ , where $\mathbf{1}$ means the singlet representation of $F_4$. Similar rule applies to other composite operators carrying a certain representation of $F_4$) is
\bea
&&\gamma_{\phi^2\in\mathbf{1}}=\frac{T_2}{2}g^2 +\frac{1}{48}  T_2 \left(24 T_3-T_2\right)g^4\nn\\
&&-\Big(432 T_3 T_2 \zeta (3)-864 T_3^2 \zeta (3)-380 T_2^2+711 T_3 T_2-1170 T_3^2-756 T_5\Big)\frac{ T_2}{1728}g^6+O(g^8)\,.\nn\\
\eea
In the above expressions, the constants $\{T_2,T_3,T_5,T_{71},T_{72}\}$ are defined as \cite{Gracey:2015tta}
\bea
&& d_{i_1 i_3 i_4}d_{i_2 i_3 i_4}=T_2 \delta_{i_1 i_2}\nn\\
&& d_{i i_1 i_2}d_{j i_1 i_3}d_{k i_2 i_3}=T_3 d_{ijk}\nn\\
&& d_{i i_1 i_2}d_{j i_3 i_4}d_{k i_5 i_6}d_{i_1 i_3 i_5}d_{i_2 i_4 i_6}=T_5 d_{ijk}\nn\\
&& d_{i i_1 i_2}d_{j i_3 i_4}d_{k i_5 i_6}d_{i_1 i_3 i_7}d_{i_2 i_5 i_8}d_{i_4 i_6 i_9}d_{i_7 i_8 i_9}=T_{71} d_{ijk}\nn\\
&& d_{i i_1 i_2}d_{j i_3 i_4}d_{k i_5 i_6}d_{i_1 i_3 i_7}d_{i_2 i_5 i_8}d_{i_4 i_8 i_9}d_{i_6 i_7 i_9}=T_{72} d_{ijk}\,.
\label{Tconstants}
\eea
Using $\gamma_{\phi}$ and $\gamma_{\phi^2\in\mathbf{1}}$, the critical exponents $\eta$ and $\nu$ can be computed via
\be
\eta= 2\gamma_{\phi}(g_*),\quad
\nu^{-1}-2+\eta= 2\gamma_{\phi^2\in\mathbf{1}}(g_*)\,.
\ee
\section{Renormalization of $F_4$ invariant $\phi^3$ theory in $D=6-2\epsilon$}\label{renormalization}
\subsection{3-Loop Renormalization of $F_4$ invariant theory}\label{F4loop}
\begin{figure}[h]
\centering
\includegraphics[scale=0.7]{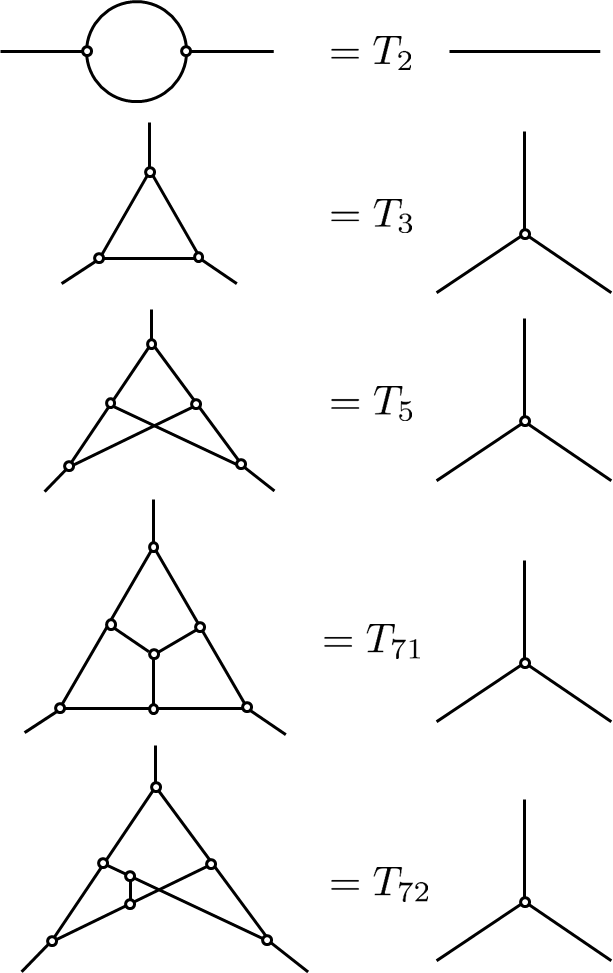}
\caption{Birdtrack diagrams defining constants $\{T_2,T_3,T_5,T_{71},T_{72}\}$.}
\label{diagram}
\end{figure}

The identities in \eqref{Tconstants} can be represented by the {\it Birdtrack} \cite{Cvitanovic:2008zz} diagrams as shown in Figure \ref{diagram}.
Some formulas used here can be found in \cite{Cvitanovic:2008zz} (see Chapter 16). The  $\mathbf{26}$  representation of $F_4$ group has the following properties:
\begin{itemize}
\item There exists a symmetric invariant rank-2 tensor $\delta_{ij}$  and a totally symmetric invariant rank-3 tensor $d_{ijk}$ carrying indices in this representation;
\item Higher rank invariant tensors carrying index in this representation are decomposable in terms of products of $\delta_{ij}$ and $d_{ijk}$ using tree diagrams;
\item The symmetric invariant rank-3 tensor $d_{ijk}$ satisfies \cite{Cvitanovic:1976am}:
\bea\label{F4defination}
&\imineq{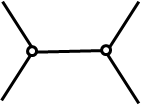}{10}+\imineq{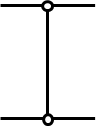}{10}+\imineq{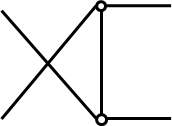}{10}=\frac{2\alpha}{n+2}\bigg(\imineq{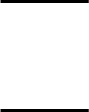}{10}+\imineq{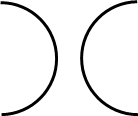}{10}+\imineq{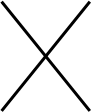}{10}\bigg)&\nn\\&
d_{il m}d_{mjk}+d_{ijm}d_{mkl}+d_{ikm}d_{mjl}=\frac{2\alpha}{n+2}(\delta_{ij}\delta_{kl}+\delta_{il}\delta_{jk}+\delta_{ik}\delta_{jl}),&
\eea
where the indices $\{i,j,k,\dots\}$ range from 1 to $n$. This relation holds for all the represenations belonging to the $F_4$ family (See Table 3). In particular, for $F_4$, $n=26$.
\end{itemize}
The normalization of $d_{ijk}$ is represented by
\bea\label{F4normailzation}
\qquad\qquad\qquad\imineq{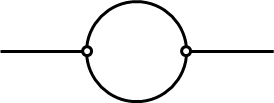}{10}&=&\alpha\imineq{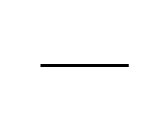}{10},\nn\\
d_{ikl}d_{jkl}&=&\alpha\delta_{ij}\,.
\eea
From now on we will set $\alpha=1$.
Notice
\be\label{F4vanishing}
\imineq{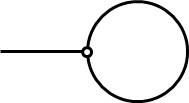}{10}=d_{imm}=0\,,
\ee
since otherwise there would exist an invariant vector $v_i=d_{imm}$.

Contracting \eqref{F4defination} with $\imineq{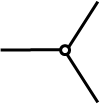}{6}$ from the left and applying \eqref{F4normailzation} and \eqref{F4vanishing}, we get
\be
\imineq{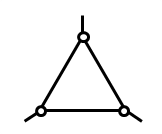}{11}=-\frac{1}{2}\frac{n-2}{n+2}\imineq{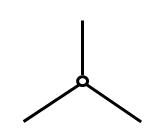}{11}.
\ee
Contracting \eqref{F4defination} with $\imineq{F4_3.png}{6}$ from the right, one gets
\be\label{F4sym}
\imineq{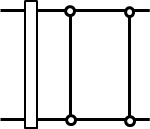}{10}=\frac{n-2}{4(n+2)}\imineq{F4_2.png}{10}+\frac{2}{n+2}\imineq{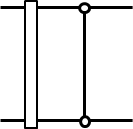}{10}+\frac{1}{n+2}\imineq{F4_6.png}{10},
\ee
where the empty box in the Birdtrack diagram means symmetrization of the two external indices. On the other hand, the antisymmetrization is given by
\bea\label{F4anti}
\imineq{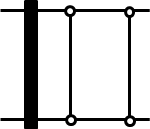}{10}&=&-\imineq{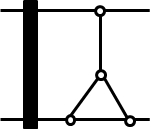}{10}-\imineq{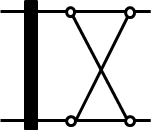}{10}+\frac{2}{n+2}\bigg(\imineq{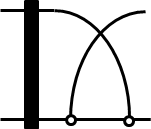}{10}+\imineq{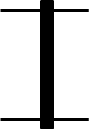}{10}\bigg)\nn\\
&=&\frac{n-6}{2(n+2)}\imineq{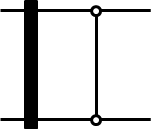}{10}+\frac{2}{n+2}\imineq{F4_18.png}{10}.
\eea
In the first step, we have used \eqref{F4defination} to replace the top $d_{ijm}d_{mkl}$ pair, and in the second step, we used the fact that the diagram $\imineq{F4_16.png}{8}$ vanishes because it is symmetric with respect to the two open indices on the left.
Combining \eqref{F4sym} and \eqref{F4anti}, we obtain
\bea
\imineq{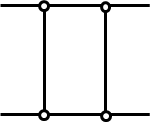}{10} =&&\frac{n-2}{4(n+2)}\bigg(\imineq{F4_2.png}{10}+\imineq{F4_3.png}{10}\bigg)+\frac{10-n}{4(n+2)}\bigg(\imineq{F4_4.png}{10}\bigg)\nn\\
&&+\frac{1}{n+2}\bigg(\imineq{F4_5.png}{10}+\imineq{F4_6.png}{10}\bigg)-\frac{1}{n+2}\imineq{F4_7.png}{10}.
\eea
Notice this equation reduces the number of vertices by two, thus one can apply such an equality to calculate all the $T$-constants. The results are given by
\bea
&& T_2 = 1\,,\quad T_3 = -\frac{n-2}{2 (n+2)}\,,\quad T_5 = -\frac{n^2-10 n-16}{2 (n+2)^2}\,,\nn\\
&& T_{71} = \frac{n^3-3 n^2+80 n+100}{4 (n+2)^3}\,,\quad T_{72} = -\frac{n \left(n^2-12 n+20\right)}{8 (n+2)^3}\,.
\eea
Substituting the values of $T$-constants to \eqref{betafunction}, we obtain the beta function up to three loops
\bea
\beta(g)&=&-\frac{\epsilon}{2} g+\frac{3 n-2}{8 (n+2)}g^3-\frac{35 n^2+296 n+596}{288 (n+2)^2}g^5\nn\\
&&+\frac{g^7}{41472 (n+2)^3}\bigg(n^3 (22032 \zeta (3)-22213)-30 n^2 (8640 \zeta (3)-9377)\nn\\&& +36 n (4464 \zeta (3)-10919)+373248 \zeta (3)-841496\bigg)\,,
\eea
which implies a unitary fixed point resides at
\bea
g_*^2&=&\frac{4 (n+2) \epsilon }{3 n-2}+\frac{4 (n+2) \left(35 n^2+296 n+596\right) \epsilon ^2}{9 (3 n-2)^3}\nn\\
&&-\frac{(n+2) \epsilon ^3}{81 (3 n-2)^5} \bigg(66096 n^4 \zeta (3)-76439 n^4-821664 n^3 \zeta (3)+722596 n^3+1000512 n^2 \zeta (3)\nn\\
&&-2776560 n^2+798336 n \zeta (3)-4560976 n-746496 \zeta (3)-1158736\bigg)\,.
\eea
At the fixed point $g=g_*$, the anomalous dimension of $\phi^i$ is given by
\bea\label{gammaphi}
\gamma_{\phi}&=&\frac{(n+2) \epsilon }{3 (3 n-2)}-\frac{2 (n+2) \left(17 n^2-174 n-296\right) \epsilon ^2}{27 (3 n-2)^3}\nn\\
&&-\frac{2 (n+2) \epsilon ^3}{243 (3 n-2)^5} \bigg(6804 n^4 \zeta (3)-8185 n^4-86184 n^3 \zeta (3)+84555 n^3+128304 n^2 \zeta (3)\nn\\&&-328018 n^2+75168 n \zeta (3)-557100 n-82944 \zeta (3)-160552\bigg)\,.
\eea
The scaling dimension of $\mathcal{O}\sim\phi^i\phi^i$ is given as
\be
\Delta_{\phi^2\in\mathbf{1}}=D-2+2\gamma_{\phi}(g_*)-2\gamma_{\phi^2\in\mathbf{1}}(g_*)\,,
\label{gammaphi21}
\ee
where
\bea
\label{gammaphi22}
&&\gamma_{\phi^2\in \mathbf{1}}=\frac{2 (n+2) \epsilon }{3 n-2}-\frac{(n+2) \left(47 n^2-868 n-1060\right) \epsilon ^2}{9 (3 n-2)^3}\nn\\
&&+\frac{\epsilon ^3}{162 (3 n-2)^5} \bigg(95159 n^5-455946 n^4+1626712 n^3+10473520 n^2\\
&&-1296 \zeta (3)(n+2) (3 n-2)(11 n^3- 184 n^2 + 116 n +288 )+10289712 n+2039264\bigg)\,,\nn
\eea
and the scaling dimension of $\mathcal{O}\sim d_{ijk}\phi^i\phi^j\phi^k$ (which we denote as $\phi^3\in \mathbf{1}$) is given by
\bea
&&\Delta_{\phi^3\in \mathbf{1}}=D+2\frac{\partial\beta}{\partial g}\Big|_{g=g_*}\,,\nn\\
&&\frac{\partial\beta}{\partial g}\Big|_{g=g_*}= \epsilon -\frac{(n (35 n+296)+596) \epsilon ^2}{9 (2-3 n)^2}\nn\\
&&+\frac{\epsilon ^3}{162 (2-3 n)^4} \bigg(-71539 n^4+805476 n^3-2259216 n^2\nn
\\&& +1296\zeta (3)(3 n-2) (17 n^3 - 200 n^2+124 n +288)-3149648 n+262128\bigg)\,.
\label{dimensionphi3}
\eea
Taking $n=26$, we have
\bea\label{anadimensionforphi2}
\Delta_{\phi}&=&\frac{D-2}2+0.12281 \epsilon-0.03152 \epsilon ^2+0.04248 \epsilon ^3+O(\epsilon^4)\,,\nn\\
\Delta_{\phi^2\in \mathbf{1}}&=&D-2-1.22807 \epsilon+0.05239 \epsilon ^2-3.41427 \epsilon ^3+O(\epsilon^4)\,,\\
\Delta_{\phi^3\in \mathbf{1}}&=&D+2\epsilon-1.22930  \epsilon ^2-0.13273\epsilon ^3+O(\epsilon^4)\,.\nn
\eea
One can see that at $D=5$ ($\epsilon=1/2$) the scaling dimensions of $\phi^i$ and $\mathcal{O}\sim d_{ijk}\phi^i\phi^j\phi^k$ receive decreasing higher loop contributions, indicating that the $F_4$ invariant fixed point may exist at $D=5$. The fact that the anomalous dimension of $\phi^i$ is positive is compatible with the unitarity of the fixed point.

Finally, we can use Pad\'{e} approximation to resum these results
\be
\text{Pad\'{e}}_{[m,n]}=\frac{A_0+A_1\epsilon+\ldots A_{m}\epsilon^m}{B_0+B_1\epsilon+\ldots B_{n}\epsilon^n}\,,
\ee
where the coefficients of $A_i$ and $B_i$ are fixed by demanding that the Taylor expansion of Pad\'{e} approximation agrees with the loop expansion. For series up to $O(\epsilon^3)$, we have two choices -- $\text{Pad\'{e}}_{[2,1]}$ and $\text{Pad\'{e}}_{[1,2]}$. We will use choose $\text{Pad\'{e}}_{[2,1]}$ to estimate \eqref{anadimensionforphi2} in the following section, because the other choice gives rise to a negative $\Delta_{\phi^2\in \mathbf{1}}$ at $D=5$. We will also provide 1-loop result when necessary.
\subsection{1-Loop Renormalization of $\phi^i\times\phi^j$ operators in $\mathbf{26}$ and $\mathbf{324}$ representations of $F_4$}\label{F4oneloop}
\begin{figure}[h]
\centering
\includegraphics[scale=0.8]{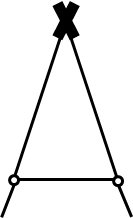}\qquad\qquad\qquad
\includegraphics[scale=0.8]{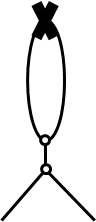}
\caption{Renormalization of operator $O^{ij}\sim \phi^i\phi^j$.}
\label{massrnl}
\end{figure}

In this section, we shall compute the 1-loop anomalous dimensions of $\phi^2$ operators transforming nontrivially under $F_4$. Minimal subtraction scheme is adopted in the calculation. We need the following projectors of $F_4$ $(n=26)$ \cite{Cvitanovic:2008zz}
\bea\label{projectorF4}
\mathbf{P}^{(\mathbf{1})}_{ijkl}&=&\frac{1}{n}\delta_{ij}\delta_{kl}\,,\nn\\
\mathbf{P}^{(\mathbf{26})}_{ijkl}&=& d_{ijm}d_{klm}\,,\nn\\
\mathbf{P}^{(\mathbf{324})}_{ijkl}&=&\frac{1}{2}\delta_{il}\delta_{jk}+\frac{1}{2}\delta_{ik}\delta_{jl}-d_{ijm}d_{klm}-\frac{1}{n}\delta_{ij}\delta_{kl}\,,\nn\\
\mathbf{P}^{(\mathbf{52})}_{ijkl}&=&\frac{8}{n+10}\left(\frac{1}{2}\delta_{il}\delta_{jk}-\frac{1}{2}\delta_{ik}\delta_{jl}+\frac{n+2}{8}(d_{ilm}d_{jkm}-d_{jlm}d_{ikm})\right)\,,\nn\\
\mathbf{P}^{(\mathbf{273})}_{ijkl}&=&\frac{n+2}{n+10}\left(\frac{1}{2}\delta_{il}\delta_{jk}-\frac{1}{2}\delta_{ik}\delta_{jl}-(d_{ilm}d_{jkm}-d_{jlm}d_{ikm})\right)\,,
\eea
which will also be useful in later conformal bootstrap study. Using these projectors, one can decompose the product representation $\mathbf{26}\times\mathbf{26}$ into the irreducible representations listed above. We will now calculate the one-loop anomalous dimensions of operators
\be
\mathcal{O}^{(I)}\sim \mathbf{P}^{(I)}_{ijkl}\phi^k\phi^l,\quad I\in \{\mathbf{1,26,324}\}\,.
\ee
These are the only ones which appear in $\phi^i(x)\times\phi^j(x)$.
In order to do so, we need to compute the three point function of the form $\langle\phi^i(p)\phi^j(q) \mathcal{O}^{(I)}(p+q)\rangle$, which receives contributions from the two Feynman diagrams in Figure \ref{massrnl}.

Using Feynman rules, these two diagrams are transferred into
\bea
D_1&=&(-\widetilde{g})^2\mathbf{P}^{(I)}_{klpq}d_{pmi}d_{qmj}{\cal I}_1=A_I \widetilde{g}^2 \mathbf{P}^{(I)}_{ijkl} {\cal I}_1\,,\qquad {\cal I}_1=\int\frac{d^D q}{(2\pi)^D}\frac{1}{(p+q)^2}\frac{1}{q^2}\,,\\
D_2&=&(-\widetilde{g})^2\mathbf{P}^{(I)}_{ijpq}d_{pqm}d_{mkl}{\cal I}_2=-B_I \widetilde{g}^2 \mathbf{P}^{(I)}_{ijkl} {\cal I}_2\,,\quad {\cal I}_2=\int\frac{d^D k}{(2\pi)^D}\frac{1}{(p-k)^2}\frac{1}{(k+q)^2}\frac{1}{k^2}\,.\nn
\eea
The index $I$ is not summed in the equations above.
Integrals ${\cal I}_1$ and ${\cal I}_2$ have been evaluated in Appendix A of \cite{Fei:2014yja}. Notice that $\widetilde{g}^2=(4\pi)^3g^2$, with the numeric factor $\text{Area}(S_5)/(2\pi)^6$ absorbed in $\widetilde{g}^2$.
In our case, the $1/\epsilon$ pole is canceled by counterterms with the coefficients
\be
\delta_{\phi^2\in I}=-A_I \frac{\widetilde{g}^2}{2 (4\pi)^3}\frac{\Gamma(\epsilon/2)}{(M^2)^{\epsilon/2}}+B_I \frac{\widetilde{g}^2}{12 (4\pi)^3}\frac{\Gamma(\epsilon/2)}{(M^2)^{\epsilon/2}}\,,
\ee
from which
\be
\gamma_{\phi^2\in I}=\frac{1}{2} M\frac{\partial}{\partial M}\delta_{\phi^2}=(\frac{A_I}{2}-\frac{B_I}{12})\frac{\widetilde{g}^2}{(4\pi)^3}=(\frac{A_I}{2}-\frac{B_I}{12})g^2\,.
\ee
After some calculations, one can check the following relations hold
\be
\mathbf{P}^{(I)}_{klpq}d_{pmi}d_{qmj}= A_I\cdot \mathbf{P}^{(I)}_{ijkl}\,, \quad
\mathbf{P}^{(I)}_{ijpq}d_{pqm}d_{mkl}= B_I \cdot \mathbf{P}^{(I)}_{ijkl}\,,
\ee
where the index $I$ is not summed, and
\bea
A_{1}&=& 1\,,\quad A_{26}=-\frac{n-2}{2 (n+2)}\,,\quad A_{324}=\frac{2}{n+2}\,,\nn\\
B_{1}&=& 0\,,\quad B_{26}=1,\qquad B_{324}=0\,.
\eea
The scaling dimensions of various operators quadratic in $\phi$ can then be computed from
\be
\Delta_{\cal O}=D-2+2\gamma_{\phi}(g_*)-2\gamma_{\cal O}(g_*)\,,
\ee
where $\gamma_{\phi}(g_*)$ is given in \eqref{gammaphi} and
\bea
\label{gamma3O}
\gamma_{\phi^2\in\mathbf{1}}&=&\frac{1}{2} M\frac{\partial}{\partial M}\delta_{\phi^2\in\mathbf{1}}\Big|_{g=g_*}=\frac{1}{2}g^2_*\Big|_{n=26}=\frac{2 (n+2) \epsilon }{3 n-2}\Big|_{n=26}+O(\epsilon^2)\,,\nn\\
\gamma_{\phi^2\in\mathbf{26}}&=& \frac{1}{2}M\frac{\partial}{\partial M}\delta_{\phi^2\in\mathbf{26}}\Big|_{g=g_*}=\frac{1-n}{3 n+6}g^2_*\Big|_{n=26}=\frac{4 (n-1) \epsilon }{6-9 n}\Big|_{n=26}+O(\epsilon^2)\,,\\
\gamma_{\phi^2\in\mathbf{324}}&=&\frac{1}{2} M\frac{\partial}{\partial M}\delta_{\phi^2\in\mathbf{324}}\Big|_{g=g_*}=\frac{1}{n+2}g^2_*\Big|_{n=26}=\frac{4 \epsilon }{3 n-2}\Big|_{n=26}+O(\epsilon^2)\,.\nn
\eea
One can see that at 1-loop level, the anomalous dimension of $\mathcal{O}^{(\mathbf{1})}\sim\phi^i\phi^i$ agrees with the result in \eqref{anadimension}.

For later comparison with the conformal bootstrap results, we summarize the scaling dimensions of various operators in Tables \ref{5dimension1} and \ref{5dimension2}.
\begin{table}[h]
\centering
\begin{tabular}{|c|c|c|c|c|c|}
 \hline \hline
&  $\Delta_{\phi}$ & $\Delta_{\phi^2\in{\mathbf{1}}}$ & $\Delta_{\phi^3\in{\mathbf{1}}}$ & $\Delta_{\phi^2\in{\mathbf{26}}}$ & $\Delta_{\phi^2\in{\mathbf{324}}}$ \\\hline
$D=5.95$ & 1.97807 & 3.91930 & 6 & 3.97807 & 3.95351 \\
\hline
$D=5$ & 1.56141 & 2.38597 & 6 & 3.56141 & 3.07018 \\
\hline
\end{tabular}
\caption{Scaling dimensions of operators in $D=5.95$ and $D=5$ at 1-loop.}\label{5dimension1}
\end{table}
\begin{table}[h]
\centering
\begin{tabular}{|c|c|c|c|}
 \hline \hline
&  $\Delta_{\phi}$ & $\Delta_{\phi^2\in{\mathbf{1}}}$ & $\Delta_{\phi^3\in{\mathbf{1}}}$ \\
\hline
$D=5.95$ & 1.97805 & 3.91931 & 5.99961  \\
\hline
$D=5$ & 1.55670 & 2.38635 & 5.83757  \\
\hline
\end{tabular}
\caption{Scaling dimensions of operators in $D=5.95$ and $D=5$ obtained from Pad\'{e}$_{2,1}$ resummed 3-loop results.}\label{5dimension2}
\end{table}

Notice that as a consequence of the equation of motion
\be
\Box \phi^i=\frac{g}{2}d_{ijk}\phi^j\phi^k\,,
\ee
the operator $\mathcal{O}^{(\mathbf{26})}\sim \mathbf{P}^{(\mathbf{26})}_{ijkl}\phi^k\phi^l$ becomes a conformal descendant operator of $\phi^i$.
From \eqref{gammaphi}, \eqref{gamma3O} and also Table \ref{5dimension1}, one can see explicitly that $\Delta_{\phi^2\in{\mathbf{26}}}=\Delta_{\phi}+2$ at 1-loop, which should also hold at higher loop level.

\section{Conformal Bootstrap}\label{conformalbootstrap}
In conformal field theories, the structures of two and three point functions are completely fixed by conformal symmetry. A four point function in CFT with $F_4$ global symmetry can be decomposed as
\bea\label{F4equations}
\langle
\contraction{}{\phi_i(x_1)}{}{\phi_j(x_2)}
\phi_i(x_1)\phi_j(x_2)
\contraction{}{\phi_k(x_3)}{}{\phi_l(x_4)}
\phi_k(x_3)\phi_{l}(x_4)
\rangle&=& \frac{1}{x_{12}^{2\Delta_{\phi}}x_{34}^{2\Delta_{\phi}}}\sum_I \mathbf{P}^{(I)}_{ijkl}\left(\sum_{{\cal O}\in I}\lambda_{\cal O}^2 g_{\Delta_{\cal O},\ell_{\cal O}}(u,v)\right)\nn\\
\text{with}\quad  I&\in&\{\mathbf{1}^+,\mathbf{26}^+,\mathbf{324}^+,\mathbf{52}^-,\mathbf{273}^-\}\ ,
\eea
where the projectors are defined in \eqref{projectorF4}. The summation runs over all conformal primary operators which appear in the OPE of $\phi_i\times \phi_j$. The function $g_{\Delta,\ell}(u,v)$ is the so called conformal block, which depends on the cross ratios
\be
u\equiv\frac{x_{12}^2 x_{34}^2}{x_{13}^2 x_{24}^2}\,,\quad v\equiv\frac{x_{14}^2 x_{23}^2}{x_{13}^2 x_{24}^2}\,,
\ee
and is completely fixed by conformal symmetry. In $D=4$, the conformal block was first obtained by Dolan and Osborn in \cite{Dolan:2000ut,Dolan:2003hv}. In other dimensions, the construction can be found in \cite{Kos:2013tga}.

Conformal bootstrap approach relies on the fact that operator algebra obeys associativity, hence the following two ways of computing four point function should lead to equivalent result
\be\label{crossingsym}
\langle
\contraction{}{\phi_i(x_1)}{}{\phi_j(x_2)}
\phi_i(x_1)\phi_j(x_2)
\contraction{}{\phi_k(x_3)}{}{\phi_{l}(x_4)}
\phi_k(x_3)\phi_{l}(x_4)
\rangle=\langle
\contraction{}{\phi_i(x_1)}{\phi_j(x_2)\phi_k(x_3)}{\phi_{l}(x_4)}
\contraction[2ex]{\phi_i(x_1)}{\phi_i(x_2)}{}{\phi_{l}(x_3)}
\phi_i(x_1)\phi_j(x_2)\phi_k(x_3)\phi_{l}(x_4)
\rangle.
\ee
This equality is also known as crossing symmetry of four point functions. Notice that the right hand side of \eqref{crossingsym} is identical to the left hand side upon the replacement $\{i,x_1\rightarrow j,x_3\}$. Initially, the four index tensor $d_{ijm}d_{kl m}$ appearing in \eqref{F4equations} and hence in \eqref{crossingsym} admits three different index structures. After utilizing the relation \eqref{F4defination}, we are left with two independent tensors $d_{jl m}d_{ikm}$ and $d_{ijm}d_{kl m}-d_{kjm}d_{il m}$,
which have definite parities (ignoring the $\delta$'s) under $i\leftrightarrow k$. Making the replacement
\bea
d_{ijm}d_{kl m}&\rightarrow&\frac{1}{2}(d_{ijm}d_{kl m}-d_{kjm}d_{il m})-\frac{1}{2}d_{ikm}d_{jl m}\nn\\&&+\frac{1}{28}(\delta_{ij}\delta_{kl}+\delta_{il}\delta_{jk}+\delta_{ik}\delta_{jl})\,,\nn\\
\frac{1}{2}d_{il m}d_{jkm}-\frac{1}{2}d_{jim}d_{ikm}&\rightarrow&-\frac{1}{4}(d_{ijm}d_{kl m}-d_{kjm}d_{il m})-\frac{3}{4} d_{jl m}d_{ikm}\nn\\&&+\frac{1}{56}(\delta_{ij}\delta_{kl}+\delta_{il}\delta_{jk}+\delta_{ik}\delta_{jl})\,,
\eea
in \eqref{crossingsym} and collecting the coefficients in front of the five independent tensor structures $\delta_{ij}\delta_{kl}\,,\delta_{ik}\delta_{jl}\,,\delta_{il}\delta_{jk}$, $d_{jl m}d_{ikm}$ and $d_{ijm}d_{kl m}-d_{kjm}d_{il m}$, we arrive at the set of equations
\bea\label{crossingeqn2}
\sum_{I}\sum_{{\cal O}\in I}\lambda_{\cal O}^2\vec{V}^{(I)}_{\Delta_{\phi},\ell_{\cal O}}[\Delta_{\cal O}]=0\,,\quad \text{with}\quad  I\in\{\mathbf{1}^+,\mathbf{26}^+,\mathbf{324}^+,\mathbf{52}^-,\mathbf{273}^-\}\,,
\eea
where the $\pm$ refers to the parity under $i\leftrightarrow k$.
The $\vec{V}^{(I)}_{\Delta_{\phi},\ell_{\cal O}}[\Delta_{\cal O}]$ are given by
\begin{alignat}{2}\label{crossingequation}
\vec{V}^{(\mathbf{1}+)}_{\Delta_{\phi},\ell_{\cal O}}[\Delta_{\cal O}]&\equiv
\left(
\begin{array}{c}
 0 \\
 0 \\
 0 \\
 \frac{F}{26} \\
 -\frac{H}{26} \\
\end{array}
\right)\,,&
\vec{V}^{(\mathbf{26}+)}_{\Delta_{\phi},\ell_{\cal O}}[\Delta_{\cal O}]&\equiv
\left(
\begin{array}{c}
 \frac{H}{2} \\
 -\tiny{F} \\
 \frac{F}{7} \\
 \frac{F}{14} \\
 0 \\
\end{array}
\right)\,,
\vec{V}^{(\mathbf{273}-)}_{\Delta_{\phi},\ell_{\cal O}}[\Delta_{\cal O}]\equiv
\left(
\begin{array}{c}
 \frac{7 H}{18} \\
 \frac{7 F}{3} \\
 -\frac{5 F}{3} \\
 \frac{F}{3} \\
 \frac{7 H}{18} \\
\end{array}
\right)\,,
\nn\\[0.4em]
\vec{V}^{(\mathbf{52}-)}_{\Delta_{\phi},\ell_{\cal O}}[\Delta_{\cal O}]&\equiv
\left(
\begin{array}{c}
 -\frac{7 H}{18} \\
 -\frac{7 F }{3}  \\
 -\frac{F}{3} \\
 \frac{F}{6} \\
 \frac{H}{9} \\
\end{array}
\right)\,,&
\vec{V}^{(\mathbf{324}+)}_{\Delta_{\phi},\ell_{\cal O}}[\Delta_{\cal O}]&\equiv
\left(
\begin{array}{c}
 -\frac{H}{2} \\
 \tiny{F} \\
 \frac{13 F}{7} \\
 \frac{71 F}{182} \\
 \frac{7 H}{13} \\
\end{array}
\right)\,.
\end{alignat}
$F$ and $H$ are the shorthand notations for the convolved blocks defined as
\bea
F_{\Delta_{\cal O},\ell_{\cal O}}^{\Delta_{\phi}}(u,v)\equiv\frac{v^{\Delta_{\phi}}g_{\Delta_{\cal O},\ell_{\cal O}}(u,v)-u^{\Delta_{\phi}}g_{\Delta_{\cal O},\ell_{\cal O}}(v,u)}{u^{\Delta_{\phi}}-v^{\Delta_{\phi}}}\,,\nn\\
H_{\Delta_{\cal O},\ell_{\cal O}}^{\Delta_{\phi}}(u,v)\equiv\frac{v^{\Delta_{\phi}}g_{\Delta_{\cal O},\ell_{\cal O}}(u,v)+u^{\Delta_{\phi}}g_{\Delta_{\cal O},\ell_{\cal O}}(v,u)}{u^{\Delta_{\phi}}+v^{\Delta_{\phi}}}\,.
\eea
We will now use the above equations to explore the conformal field theory with $F_4$ global symmetry in $D=5.95$ and $D=5$. Different from the O$(N)$ conformal bootstrap, here the fundamental field $\phi^i$ appears in its own OPE due to the cubic self-interaction.

We shall assume that the second lowest scalar primary in the $\mathbf{26}^+$ channel has scaling dimension $\Delta\geq \Delta^{\rm 2nd}_{\mathbf{26}}$. (The lowest scalar primary in this channel is just $\phi^i$.) To test this assumption, we search for a linear functional $\alpha$ with the following properties

\begin{alignat}{4}
&\alpha(\vec{V}^{(\mathbf{1}+)}_{\Delta_{\phi},0}[0])=1\,,   &\qquad\qquad &\nn \\[0.35em]
&\alpha(\vec{V}^{(\mathbf{1}+)}_{\Delta_{\phi},0}[\Delta])\geq 0\,,  & &\text{for }\Delta\geq \frac{D-2}{2}\,,\nn\\[0.35em]
&\alpha(\vec{V}^{(\mathbf{1}+)}_{\Delta_{\phi},\ell}[\Delta])\geq 0\,,  & &\text{for }\Delta\geq \ell+D-2\,, (\ell=2,4,6\dots)\,,\nn\\[0.35em]
&\alpha(\vec{V}^{(\mathbf{26}+)}_{\Delta_{\phi},0}[\Delta_{\phi}])\geq0\,, &\qquad\qquad &\nn\\[0.35em]
&\alpha(\vec{V}^{(\mathbf{26}+)}_{\Delta_{\phi},0}[\Delta])\geq0\,, & &\text{for }\Delta\geq \Delta^{\rm 2nd}_{\mathbf{26}}\,,\nn\\[0.35em]
&\alpha(\vec{V}^{(\mathbf{26}+)}_{\Delta_{\phi},\ell}[\Delta])\geq 0\,, & &\text{for }\Delta\geq \ell+D-2\,, (\ell=2,4,6\dots)\,,\nn\\[0.35em]
&\alpha(\vec{V}^{(\mathbf{324}+)}_{\Delta_{\phi},0}[\Delta])\geq 0\,, & &\text{for }\Delta\geq \frac{D-2}{2}\,,\nn\\[0.35em]
&\alpha(\vec{V}^{(\mathbf{324}+)}_{\Delta_{\phi},\ell}[\Delta])\geq 0\,, & &\text{for }\Delta\geq \ell+D-2\,, (\ell=2,4,6\dots)\,,\nn\\[0.35em]
&\alpha(\vec{V}^{(\mathbf{52}-)}_{\Delta_{\phi},\ell}[\Delta])\geq 0\,,  & &\text{for }\Delta\geq \ell+D-2 \,,(\ell=1,3,5\dots)\,,\nn\\[0.35em]
&\alpha(\vec{V}^{(\mathbf{273}-)}_{\Delta_{\phi},\ell}[\Delta])\geq 0\,, & & \text{for }\Delta\geq \ell+D-2 \,,(\ell=1,3,5\dots)\,.
\end{alignat}

For a given choice of $(\Delta_{\phi},\Delta^{\rm 2nd}_{\mathbf{26}})$, if such a linear functional $\alpha$ exists, then we have
\be
\alpha\left(\sum_I\sum_{{\cal O}\in I}\lambda_{\cal O}^2\vec{V}^{(I)}_{\Delta_{\phi},\ell_{\cal O}}[\Delta_{\cal O}]\right)=\sum_I\sum_{{\cal O}\in I}\lambda_{\cal O}^2\alpha(\vec{V}^{(I)}_{\Delta_{\phi},\ell_{\cal O}}[\Delta_{\cal O}])>0\,,
\ee
which contradicts \eqref{crossingeqn2}. The contradiction simply implies that the second lowest scalar primary in the $\mathbf{26}^+$ channel should have scaling dimension lower than $\Delta^{\rm 2nd}_{\mathbf{26}}$. We use ``SDPB'' \cite{Simmons-Duffin:2015qma} to implement numerical bootstrap. For technical reason, we have to restrict the number of derivatives and the range of spins involved in the numerical calculation. The maximal derivative orders are chosen to be $\Lambda=19,21$ or $23$ according to necessity, and the corresponding ranges of spins are set to be  $\ell\in\{1,\ldots26\}\cup\{49,50\}$, $\ell\in\{1,\ldots30\}\cup\{49,50\}$ and $\ell\in\{1,\ldots30\}\cup\{49,50\}$ respectively.
\begin{figure}[h]
\centering
\includegraphics[scale=0.65]{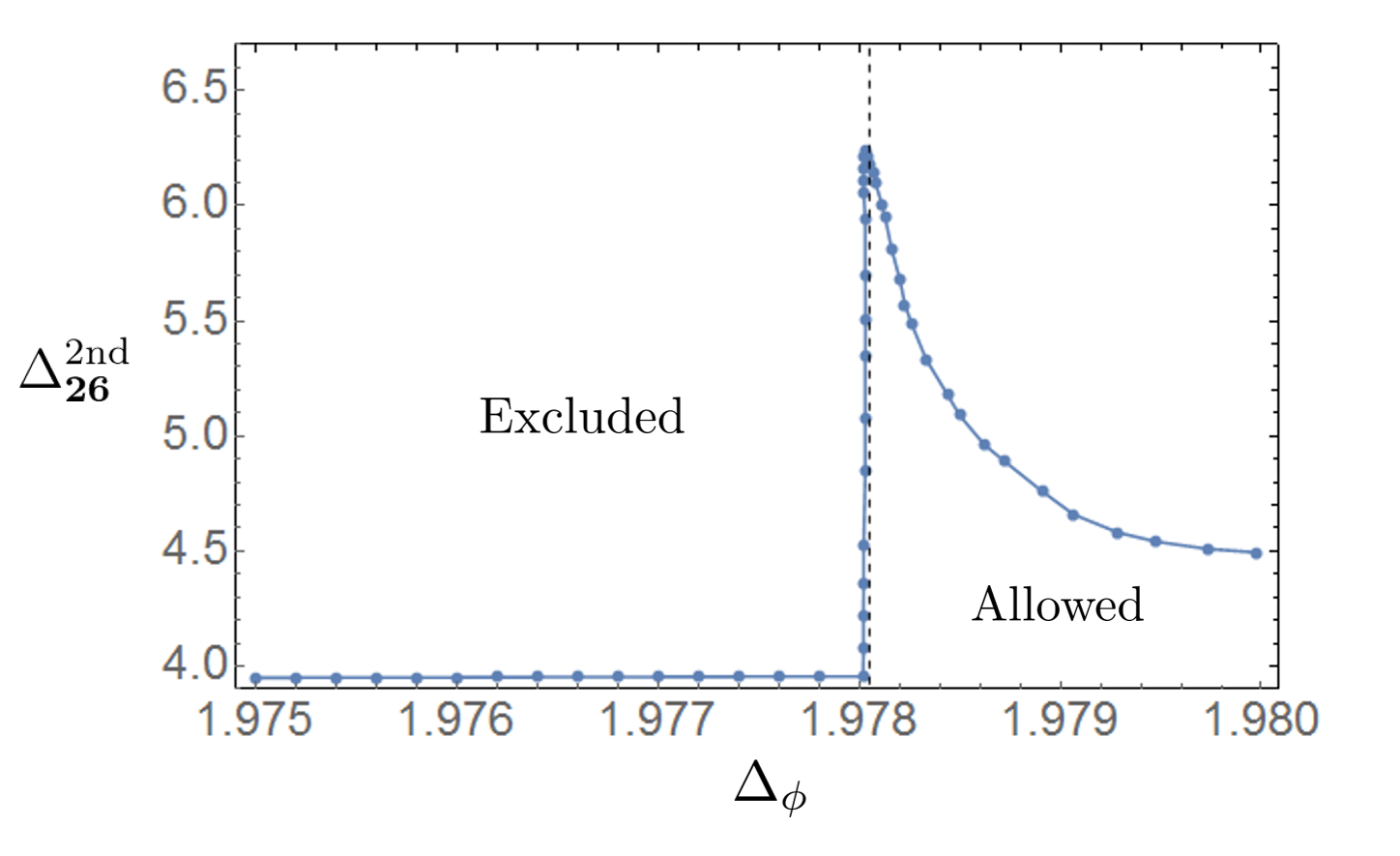}
\caption{Bootstrap study of $F_4$ invariant $\phi^3$ theory in $D=5.95$. Allowed region of $(\Delta_{\phi},\Delta_{\mathbf{26}}^{\rm 2nd})$ is indicated. The dashed line indicates $\Delta_{\phi}$ from Pad\'{e}$_{[2,1]}$ approximation of the 3-loop result. The curve is obtained at $\Lambda=19$.}
\label{5.95DF4}
\end{figure}

In $D=5.95$, the result is shown in Figure \ref{5.95DF4}. In this case, $\epsilon=0.025$ thus one expects the perturbative calculation to be valid. The boundary of the allowed region in the $(\Delta_{\phi},\Delta_{\mathbf{26}}^{\rm 2nd})$ plane exhibits a sharp peak at $\Delta_{\phi}\approx 1.978$, which coincides precisely with the value of $\Delta_{\phi}$ given by the $\text{Pad\'{e}}_{[2,1]}$ resummed 3-loop result. The is a highly nontrivial check of the perturbative calculations performed in previous sections. Since $\mathcal{O}^{(\mathbf{26})}\sim d_{ijk}\phi^i\phi^k$ is a descendant operator, one expects the dimension of the second lowest scalar primary in this channel to be much higher than $D-2$ (the classical dimension of $\mathcal{O}^{(\mathbf{26})}$), which is indeed the case.
\begin{figure}[h]
\centering
\includegraphics[scale=0.65]{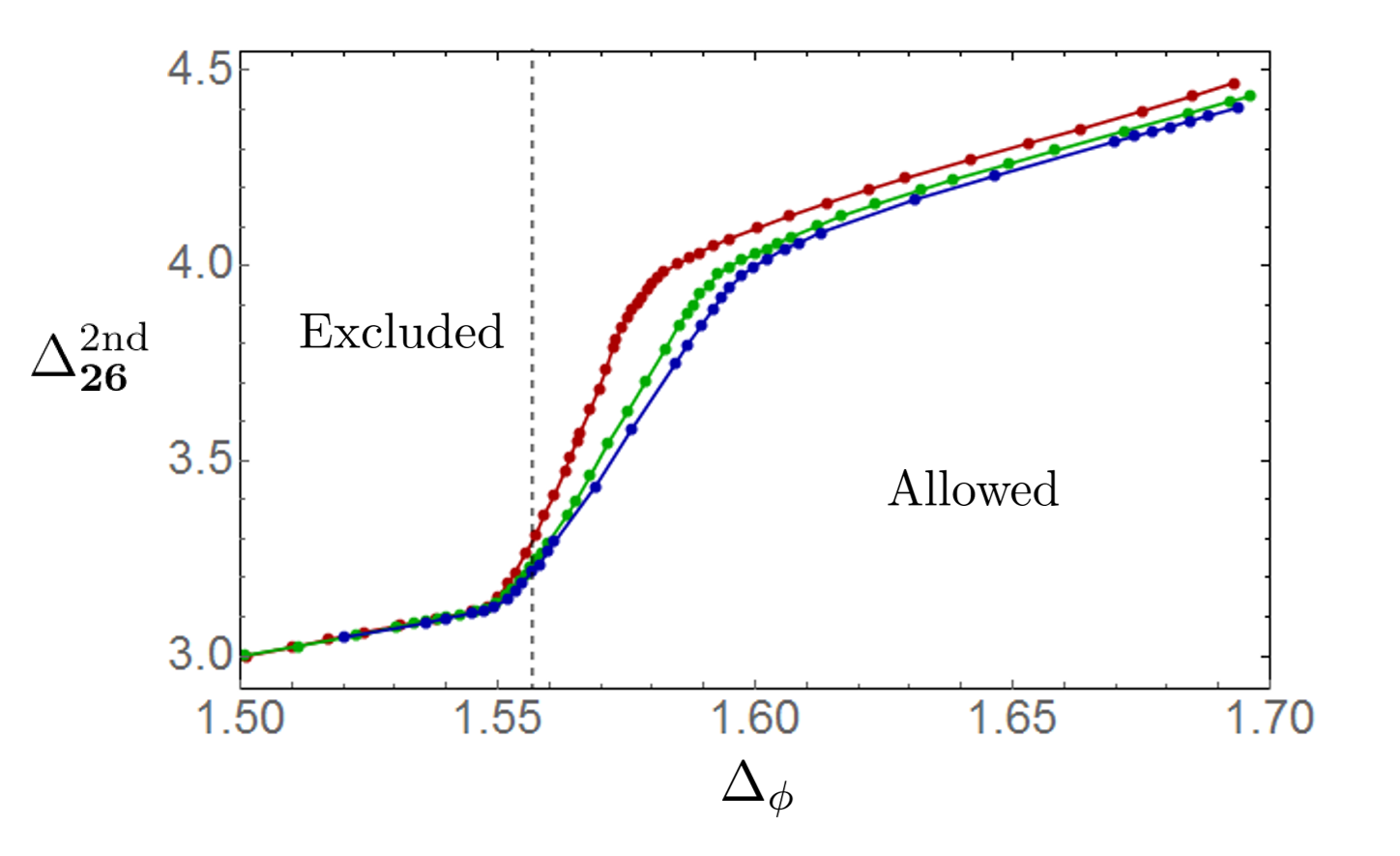}
\caption{Bootstrap study of $F_4$ invariant $\phi^3$ theory in $D=5$. Allowed region of $(\Delta_{\phi},\Delta_{\mathbf{26}}^{\rm 2nd})$ is indicated. The dashed line indicates $\Delta_{\phi}$ from Pad\'{e}$_{[2,1]}$ approximation of the 3-loop result. The red, green and blue curves correspond to $\Lambda=19,21,23$ respectively. }
\label{5DF4}
\end{figure}

The $D=5$ result is plotted in Figure \ref{5DF4}. Unlike the $D=5.95$ case, no peak seems to appear. However, a weak kink is observed at $\Delta_{\phi}\approx1.6$. The three loop result $\Delta_{\phi}\approx1.55670$ is indicated by the dashed line. The location of the week kink does not fit the 3-loop result may be expected, since in this case, $\epsilon=0.5$, and the $D=5$ fixed point is highly non-perturbative in nature. We also notice that when increasing the derivative order, the weak kinks corresponding to different derivative orders tend to converge to a single weak kink.
\section{Discussion}\label{discussion}
The weak kink observed on the $D=5$ bootstrap curve indicates the possibility of the existence of a CFT with $F_4$ flavor symmetry.  It would be interesting to further constrain the $D=5$ fixed point using mixed correlators conformal bootstrap \cite{Kos:2014bka}. We leave this for future investigation.

The exceptional Lie group of $F_4$ belongs to the so called $F_4$ family of invariant groups, which is defined as the family of groups admitting a representation that satisfies the three conditions listed in Section \ref{F4loop}. According to the Birdtrack classification, there are four choices, the groups and the dimensions of the relevant representations are listed  in Table \ref{F4table}. There exists a fixed point for each of these choices in $6-2\epsilon$ dimensions. In Appendix \ref{F4familysepcturm}, we list the dimensions of various operators computed for these theories at the fixed point. They are obtained simply by substituting the value of $n$ for each case to formulas given in Section \ref{F4loop} and Section \ref{F4oneloop}. Whether some of these fixed points continue to exist in $D=5$ or even $D<5$ is also worth further study. Another interesting question is whether other resummation method would give us a better estimation of the operator dimension.
\begin{table}[h]
\centering
\begin{tabular}{|l|c|c|c|c|}
 \hline \hline
 Group &  $F_4$ & $B_1\equiv A_1$&$A_2$ & $C_3$ \\\hline
$ \phi^i\in\mathbf{n}$ & 26 & 5 & 8 & 14\\\hline
\end{tabular}
\caption{$F_4$ family of invariant groups and the dimensions of the relevant representations. Here we use capital letter to label the compact real form of the Lie algebra.}\label{F4table}
\end{table}

It is also possible to consider more general $\phi^3$ theory of the form \eqref{F4Lagragian} which does not necessarily belongs to the $F_4$ family by relaxing one of the conditions in Section \ref{F4defination}, and then classifying flavor symmetry groups which allow a stable nontrivial IR fixed point.  For instance, $SU$(N) with N$\geq$3 possesses a rank-3 symmetric invariant tensor in the adjoint representation which satisfies the first two conditions but not the third one in Eq. \eqref{F4defination}. The four loop beta function for $SU$(N) invariant $\phi^3$ theory has been computed in \cite{Gracey:2015tta}. It would be interesting to carve out the possible IR fixed points in $ D<6$  using conformal bootstrap approach.

Besides the $F_4$ family, there is also the so called $E_6$ family of invariant groups \cite{Cvitanovic:2008zz}. The groups admits an invariant 2-tensor $\delta^i_{~j}$ and an invariant symmetric 3-tensor $d_{ijk}$ (and its conjugate $d^{ijk}$) carrying indices in some representation and satisfying certain conditions similar to those listed in \ref{F4defination}. The groups and dimensions of the relevant representations are summarized in Table \ref{E6table}.
\begin{table}[h]
\centering
\begin{tabular}{|l|c|c|c|c|}
 \hline \hline
 Group &  $E_6$ & $A_5$&$A_2\times A_2$ & $A_2$ \\\hline
$ \phi^i\in\mathbf{n}$ & 27 & 15 & $3\times3$=9 & 6\\\hline
\end{tabular}
\caption{$E_6$ family of invariant groups and the dimensions of the relevant of representations. Here the capital letter labels the compact real form of the corresponding Lie algebra.}\label{E6table}
\end{table}
One can then write down the Lagrangian $$\mathcal{L}=\partial_{\mu}\phi^i\partial^{\mu}\bar{\phi}_i+\frac{g}{6}(d_{ijk}\phi^i\phi^j\phi^k+d^{ijk}\bar{\phi}_i\bar{\phi}_j\bar{\phi}_k)$$ which is invariant under the $E_6$-family of groups. The $SU(3)\times SU(3)$ invariant theory considered in \cite{McKane:1976zz,Mckane:1977bv,Gracey:2015tta} is just the special case with $n=9$. It was argued in \cite{Gracey:2015tta} that the formula \eqref{betafunction} is still applicable after one sets $T_3=T_{72}=0$ (one could check the corresponding diagrams do not exist if replacing $\imineq{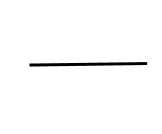}{8}$ by $\imineq{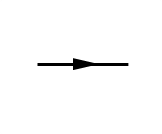}{8}$, and $\imineq{F4_11.png}{8}$ by $\imineq{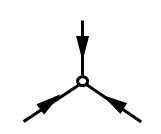}{8}$ or $\imineq{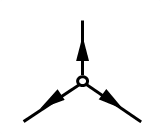}{8}$). Fixing the normalization by choosing $T_2=1$, the 1-loop beta function is given by
\be
\beta(g)=-\frac{\epsilon}{2}g+\frac{1}{8}g^2.
\ee
One can see that there exists a stable unitary fixed point with $g_{*}^2=4\epsilon$. It would be interesting to study renormalization of the $\phi^3$ theory invariant under the $E_6$ family at higher loop level using Birdtrack technique, and figure out whether the 1-loop fixed point continues to exist at larger value of $\epsilon$ using conformal bootstrap.
\section*{Acknowledgements}

We are grateful to Zhijin Li, Chris Pope, Soo-Jong Rey for helpful discussion. Y.P. would like to thank Beijing Normal University and Sun Yet-sen University for hospitality during various stages of this work. J.R. and N.S. would like to thank George P. \& Cynthia Woods Mitchell Institute for Fundamental Physics and Astronomy in Texas A\&M University, where the early stage of this work was completed. The work of Y.P. is supported by Alexander von Humboldt fellowship. The work of J.R. was supported by DOE grant DE-FG02-13ER42020. The work of N.S. was supported by NSF grants PHY-1214222 and PHY-1521099.
\appendix
\section{3-loop renormalization for $F_4$ family of invariant groups}\label{F4familysepcturm}

The $F_4$ family of groups and the corresponding representations listed in Table \ref{F4table} share the common properties mentioned in \eqref{F4defination}. Moreover, if denoting the relevant representation in each case by $ [\mathbf{n}]$, the product of two $[\mathbf{n}]$s admits similar decomposition rules which are summarized as
\be
[\mathbf{n}]\times [\mathbf{n}] \rightarrow[\mathbf{1}]_{+}+[\mathbf{\frac{3 n(n-2)}{n+10}}]_{-}+[\mathbf{\frac{ n(n+1)(n+2)}{2(n+10)}}]_{-}+[\mathbf{n}]_{+}+[\mathbf{\frac{n(n-1)}{2}}-1]_{+}\,,
\ee
where the number in the square bracket indicates the dimension of the irreducible representation and the  subscript $\pm$ refers to symmetry properties under the interchange of the two indices in the $[\mathbf{n}]$ representation. The $[\mathbf{\frac{3 n(n-2)}{n+10}}]$ representation corresponds to the adjoint representation. In this appendix, we will present the 3-loop renormalized dimensions for operators $\phi^i$, $\phi^i\phi^i$ and  $d_{ijk}\phi^i\phi^j\phi^k$ and the 1-loop renormalized dimensions for operators $\mathbf{P}^{(\mathbf{n})}_{ijkl} \phi^k\phi^l$ and $\mathbf{P}^{(\mathbf{\frac{n(n-1)}{2}-1})}_{ijkl} \phi^k\phi^l$, where the projectors $\mathbf{P}_{ijkl}$ are defined in \eqref{projectorF4}. Similar to the $F_4$ case, we will denote these operators by $\{\phi,\phi^2\in\mathbf{1},\phi^3\in\mathbf{1}\}$ and $\{\phi^2\in \mathbf{n},\phi^2\in \mathbf{\frac{n(n-1)}{2}-1}\}$ for convenience. In Section \ref{renormalization}, when discussing the renormalization of $F_4$ invariant $\phi^3$ theory, we have left $n$ unspecified in the intermediate steps. Therefore, the results for other members in the $F_4$ family can be obtained by simply substituting the value of $n$ in each case to Eqs. \eqref{gammaphi}, \eqref{gammaphi21}, \eqref{gammaphi22}, \eqref{dimensionphi3} and \eqref{gamma3O}.

In summary, for $F_4$ group case, $n=26$,
\bea
\Delta_{\phi}&=&\frac{D-2}{2}+0.12281 \epsilon-0.03152 \epsilon ^2+0.04248 \epsilon ^3+O(\epsilon^4)\,,\nn\\
\Delta_{\phi^2\in\mathbf{1}}&=& D-2-1.22807 \epsilon+0.05239 \epsilon ^2-3.41428 \epsilon ^3+O(\epsilon^4)\,,\nn\\
\Delta_{\phi^3\in\mathbf{1}}&=& 6-1.22930 \epsilon ^2-0.13273 \epsilon ^3+O(\epsilon^4)\,,\nn\\
\Delta_{\phi^2\in\mathbf{26}}&=&D-2+1.12281 \epsilon+O(\epsilon^2)\,,\nn\\
\Delta_{\phi^2\in\mathbf{324}}&=& D-2+0.14035 \epsilon+O(\epsilon^2);
\eea
for $B_1\equiv A_1$ group case, $n=5$,
\bea
\Delta_{\phi}&=&\frac{D-2}{2}+0.17949\epsilon+0.17489 \epsilon ^2+1.44664 \epsilon ^3+O(\epsilon^4)\,,\nn\\
\Delta_{\phi^2\in\mathbf{1}}&=& D-2-1.79487 \epsilon-2.64168 \epsilon ^2-25.8755 \epsilon ^3+O(\epsilon^4)\,,\nn\\
\Delta_{\phi^3\in\mathbf{1}}&=& 6-3.88034 \epsilon ^2-24.1329 \epsilon ^3+O(\epsilon^4)\,,\nn\\
\Delta_{\phi^2\in\mathbf{5}}&=&D-2+1.17949 \epsilon+O(\epsilon^2)\,,\nn\\
\Delta_{\phi^2\in\mathbf{9}}&=& D-2-0.25641 \epsilon+O(\epsilon^2)\,;
\eea
for $A_2$ group case, $n=8$,
\bea
\Delta_{\phi}&=&\frac{D-2}{2}+0.15152 \epsilon+0.041740 \epsilon ^2+0.39753 \epsilon ^3+O(\epsilon^4)\,,\nn\\
\Delta_{\phi^2\in\mathbf{1}}&=& D-2-1.51515\epsilon-0.959179 \epsilon ^2+10.0498 \epsilon ^3+O(\epsilon^4)\,,\nn\\
\Delta_{\phi^3\in\mathbf{1}}&=& 6-2.38935 \epsilon ^2-7.72911 \epsilon ^3+O(\epsilon^4)\,,\nn\\
\Delta_{\phi^2\in\mathbf{8}}&=&D-2+1.15152 \epsilon+O(\epsilon^2)\,,\nn\\
\Delta_{\phi^2\in\mathbf{27}}&=& D-2-0.06061\epsilon+O(\epsilon^2)\,;
\eea
for $C_3$ group case, $n=14$,
\bea
\Delta_{\phi}&=&\frac{D-2}{2}+0.13333 \epsilon-0.01111 \epsilon ^2+0.12005 \epsilon ^3+O(\epsilon^4)\,,\nn\\
\Delta_{\phi^2\in\mathbf{1}}&=& D-2-1.33333\epsilon-0.24444 \epsilon ^2-5.09869 \epsilon ^3+O(\epsilon^4)\,,\nn\\
\Delta_{\phi^3\in\mathbf{1}}&=& 6-1.61111\epsilon ^2-2.09496 \epsilon ^3+O(\epsilon^4)\,,\nn\\
\Delta_{\phi^2\in\mathbf{14}}&=&D-2+1.13333 \epsilon+O(\epsilon^2)\,,\nn\\
\Delta_{\phi^2\in\mathbf{90}}&=& D-2+0.06667 \epsilon+O(\epsilon^2)\,.
\eea

\end{document}